\documentclass[twocolumn,twocolappendix]{aastex63}
\usepackage{amssymb, amsfonts, amsmath,framed}
\usepackage{lineno}

\usepackage[utf8]{inputenc}
\usepackage{tabularx}
\usepackage{multirow}
\usepackage{url}
\usepackage{amsmath}
\usepackage{gensymb}
\usepackage{graphicx}
\usepackage{mathrsfs}
\usepackage{color}
\usepackage[normalem]{ulem}
\usepackage{float}
\usepackage{lmodern}
\usepackage{soul}
\usepackage{graphicx,verbatim}
\usepackage{stackengine}
\usepackage[caption=false]{subfig}
\usepackage{threeparttable}
\usepackage{csquotes}

\newcommand{\be}{\begin{equation}}
\newcommand{\ee}{\end{equation}}
\newcommand{\bi}{\begin{itemize}}
\newcommand{\ei}{\end{itemize}}

\newcommand{\fix}[1]{\textcolor{red}{[fix] }}

\definecolor{darkblue}{rgb}{0.0, 0.0, 0.5}

\newcommand{\dk}[1]{{\color{black} #1}} 

\begin{document}

\title{Guide field effects on particle acceleration in three-dimensional relativistic magnetic reconnection}

\author[0000-0002-2383-3860]{Despina Karavola}
\email{dkaravola@phys.uoa.gr}
\affiliation{Department of Physics, National and Kapodistrian University of Athens, University Campus Zografos, GR 15784, Athens, Greece}
\author[0000-0002-1227-2754]{Lorenzo Sironi}
\affiliation{Department of Astronomy and Columbia Astrophysics Laboratory, Columbia University, New York, NY 10027, USA}
\affiliation{Center for Computational Astrophysics, Flatiron Institute, 162 5th Avenue, New York, NY, 10010, USA}
\author[0000-0001-6640-0179]{Maria Petropoulou}
\affiliation{Department of Physics, National and Kapodistrian University of Athens, University Campus Zografos, GR 15784, Athens, Greece}
\affiliation{Institute of Accelerating Systems \& Applications, University Campus Zografos, Athens, Greece}


\date{\today}

\begin{abstract}
Relativistic magnetic reconnection is a key mechanism for particle acceleration in astrophysical plasmas. Recent 3D particle-in-cell simulations have shown that, for weak guide fields, particles can be accelerated at nearly the maximal rate by escaping the reconnected plasma and gaining energy from the large-scale upstream electric field. Here, we investigate whether this fast acceleration channel persists as the guide field becomes comparable to the reconnecting field. We perform 3D simulations of relativistic pair-plasma reconnection, complemented by corresponding 2D runs, using open boundaries along the reconnection outflows to establish a quasi-steady state. We find that increasing the guide field strength reduces the reconnection rate and causes energetic particles to become more closely aligned with the local magnetic field. In 3D, the fast acceleration channel---absent in 2D---persists across the full range of guide fields investigated, but involves a progressively smaller fraction of particles as the guide field increases. Consequently, stronger guide fields produce steeper particle energy spectra and lower maximum Lorentz factors.
\end{abstract}



\section{Introduction} \label{Sec:Intro}
Magnetic reconnection is a fundamental mechanism for rapid magnetic energy dissipation and efficient particle energization in space and astrophysical environments. In the relativistic regime, where the magnetic energy density greatly exceeds the particle rest-mass energy density, or equivalently the plasma magnetization is $\sigma\gg1$  \citep{lyutikov_uzdensky_03, lyubarsky_05, comisso_14, sironi_15}, reconnection can efficiently accelerate particles into nonthermal power-law distributions, potentially powering the nonthermal emission observed from a broad range of astrophysical sources (see \citet{sironi_review_25} for a recent review and references therein).

Our understanding of particle acceleration in relativistic reconnection has advanced considerably in recent years, primarily through large-scale, first-principles particle-in-cell (PIC) simulations. Most studies have employed two-dimensional (2D) configurations that resolve the plane of the reconnecting fields \citep{zenitani_01, zenitani_07, sironi_spitkovsky_14, guo_14, guo_15a, werner_16, cerutti_15, comisso_pitch-angle_2023}. More recently, large-scale three-dimensional (3D) simulations have revealed qualitatively new behavior, particularly in the mechanisms responsible for particle acceleration \citep{sironi_spitkovsky_14, werner_17, zhang_sironi_21, zhang_23, chernoglazov_23, hakobyan_reconnection-driven_2025, Deka_2026_preprint}.

Recent 3D PIC simulations of relativistic reconnection have demonstrated a fast particle-acceleration channel that is intrinsically three-dimensional \citep{zhang_sironi_21, zhang_23}. A subset of particles can escape from the plasmoids / flux ropes in the  downstream plasma and go back to the upstream region, where they meander around the current layer, repeatedly sampling the two sides of the reconnection plane. During this \enquote{free} phase, particles gain most of their energy from the large-scale ideal electric field that threads the entire reconnection volume. Their acceleration rate can approach the theoretical maximum, $\dot{\gamma}_{\rm acc} \sim \beta_{\rm rec}\,\omega_{\rm c}$, where $\omega_{\rm c}=eB_{\rm 0}/mc$ is the cyclotron frequency and \dk{$\beta_{\rm rec}$} is the reconnection rate \citep{zhang_sironi_21, zhang_23}. The free phase terminates when particles become trapped again within downstream flux ropes, where further energization proceeds much slower \citep{zhang_sironi_21}. This sequence of free acceleration and subsequent trapping produces a nonthermal power-law energy spectrum whose slope depends only weakly on the magnetization, as long as $\sigma\gtrsim \,$few \citep{zhang_23}.

So far, the physics and efficiency of this fast acceleration process have been investigated only in simulations with weak or vanishing guide fields. \dk{The term \enquote{guide field} is used to indicate} a uniform field component aligned with the electric current, or equivalently perpendicular to the plane of the reconnecting  fields. Previous studies of relativistic reconnection have shown that increasing the guide-field strength reduces the reconnection rate and the fraction of magnetic energy dissipated into particles \citep[e.g.][]{zenitani_05b, zenitani_08, melzani_14, sironi_15, werner_17, french_22, werner_21, werner_24, hoshino_24, Deka_2026_preprint}. This reduced dissipation for stronger guide fields is accompanied by a lower efficiency in nonthermal particle acceleration and by steeper particle energy spectra, a trend observed in both 2D and 3D PIC simulations \citep[e.g.,][]{werner_17, comisso_pitch-angle_2023, comisso_concurrent_2024}.

In this paper, we employ large-scale 3D PIC simulations to determine whether the fast particle-acceleration channel, intrinsically three-dimensional and so far investigated only for weak guide fields, persists in the presence of moderate or strong guide fields. We vary the guide-field strength up to values comparable to the reconnecting field. We find that the fast acceleration mechanism remains robust across the full range of guide fields investigated, but involves a progressively smaller fraction of particles as the guide field strength increases.

Our simulations employ outflow boundary conditions along the reconnection exhausts, allowing the system to attain a quasi-steady state. We confirm that increasing the guide-field strength reduces the reconnection rate and steepens the overall particle energy spectrum. At each guide-field strength, we compare the 3D simulations with corresponding 2D runs in order to isolate genuinely 3D effects. We also measure the pitch-angle anisotropy of the accelerated particles as a function of energy and guide-field strength, extending the results of earlier 2D studies \citep{comisso_pitch-angle_2023, comisso_concurrent_2024} to three dimensions. 

This paper is organized as follows. In Section~\ref{sec:Setup}, we describe the numerical setup of our PIC simulations. In Section~\ref{Sec:results}, we present our main results as a function of guide-field strength, including the structure of the reconnection layer, the reconnection rate, particle energy spectra, and the acceleration physics of the highest-energy particles. In Section~\ref{Sec:summary}, we summarize our main findings and discuss their astrophysical implications.

 
\section{Simulation Setup}\label{sec:Setup}

We perform 2D and 3D PIC simulations employing the TRISTAN-MP code \citep[][]{buneman_93, spitkovsky_05}. We initialize a Harris equilibrium with  reconnecting magnetic field that reverses from $+ {\hat x}$ to $-{\hat x}$ at $y=0$. Specifically, $\mathbf{B}= B_{\rm 0} \tanh(y/\Delta) \,{\hat x}$, with $\Delta$ being the half-thickness of the current layer. The magnetic field strength $B_{\rm 0}$ is parametrized by the magnetization $\sigma =B_{\rm 0}^2/4\pi n_0mc^2 = \left(\omega_{\rm  c} / \omega_{\rm  p}\right)^2$, where $\omega_{\rm  c} = e B_{\rm 0} / m c$ and $\omega_{\rm  p} = \sqrt{4\pi n_0 e^2 / m}$ are the gyrofrequency and the plasma frequency, respectively. Here, $n_0$ is the number density of the upstream electron-positron plasma, having particle mass $m$ and elementary charge $e$. We fix $\sigma=10$ as a representative choice for relativistic reconnection. In addition to the reconnecting field, we initialize a uniform magnetic field component along the $z$ direction (i.e., the guide field) of strength $B_{\rm g}$. We vary the guide field strength in the range  $B_{\rm g}/B_0=\{0, 0.1, 0.3, 0.5, 1.0\}$. 

We initialize the 3D simulation domain with an average of two macroparticles per cell, including both species (\citealt{zhang_sironi_21} explored a larger number of particles per cell and found no significant changes). The 2D simulations contain 32 macroparticles per cell and are otherwise initialized with numerical and physical parameters identical to those of the corresponding 3D runs. In both 2D and 3D, we resolve the plasma skin depth $c/\omega_{\rm p}$ with 2.5 cells; the speed of light is 0.45 cells/timestep.

We employ periodic boundary conditions along the $z$ direction, which is aligned with both the electric current and the guide field. Along the $y$ direction of the reconnection inflows, two injectors continuously introduce fresh plasma and magnetic flux into the domain while moving away from the midplane. Along the $x$ direction, corresponding to the reconnection outflows, we employ outflow boundary conditions \citep{sironi_16}. In our analysis of both 2D and 3D simulations, we exclude 200 cells (equivalently, $80\, c/\omega_{\rm p}$) from each end of the domain along $x$ in order to minimize potential artifacts induced by the outflow boundaries.

\begin{figure*}
    \centering
    \includegraphics[width=0.49\linewidth]{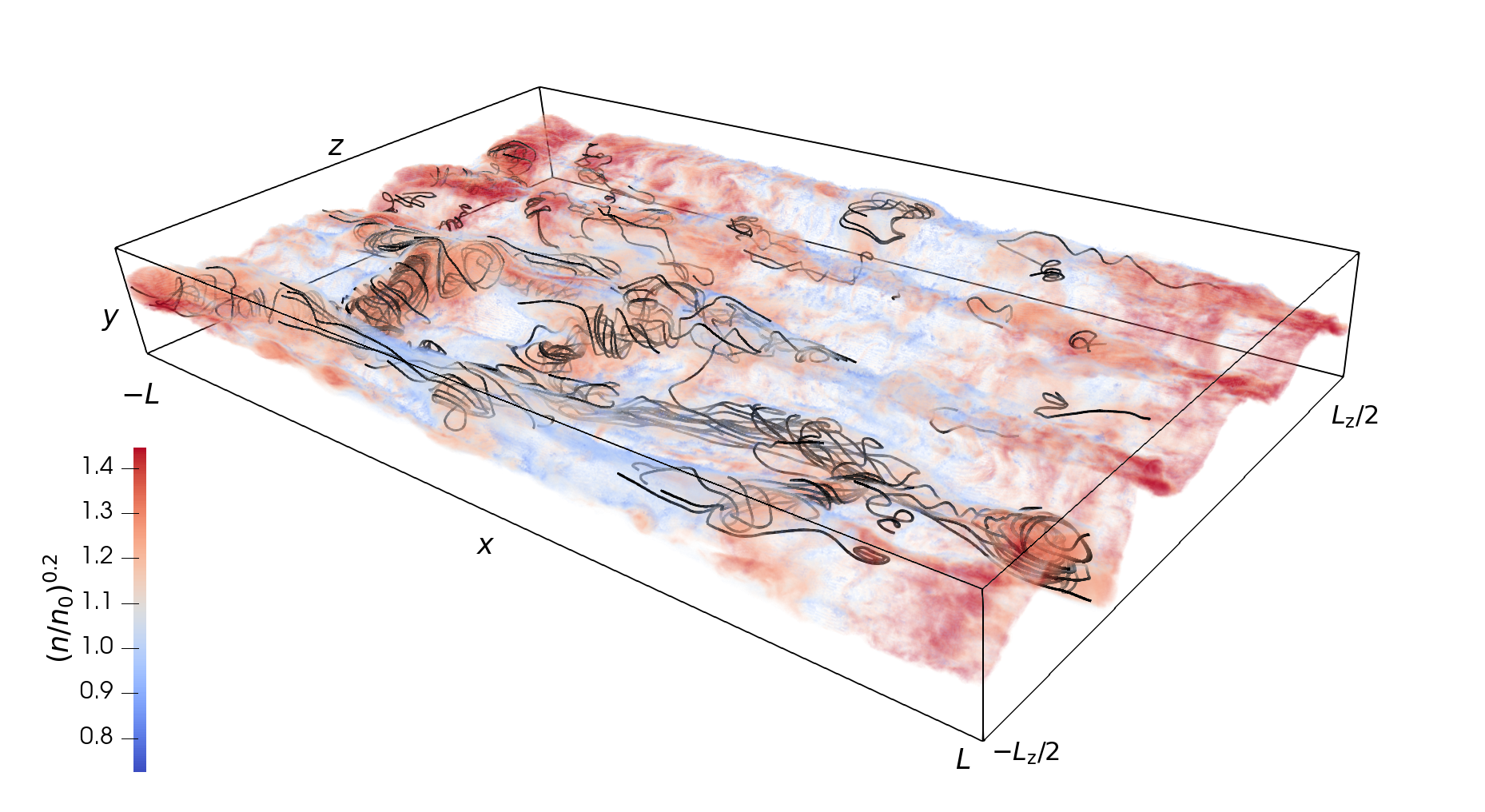}
    \includegraphics[width=0.49\linewidth]{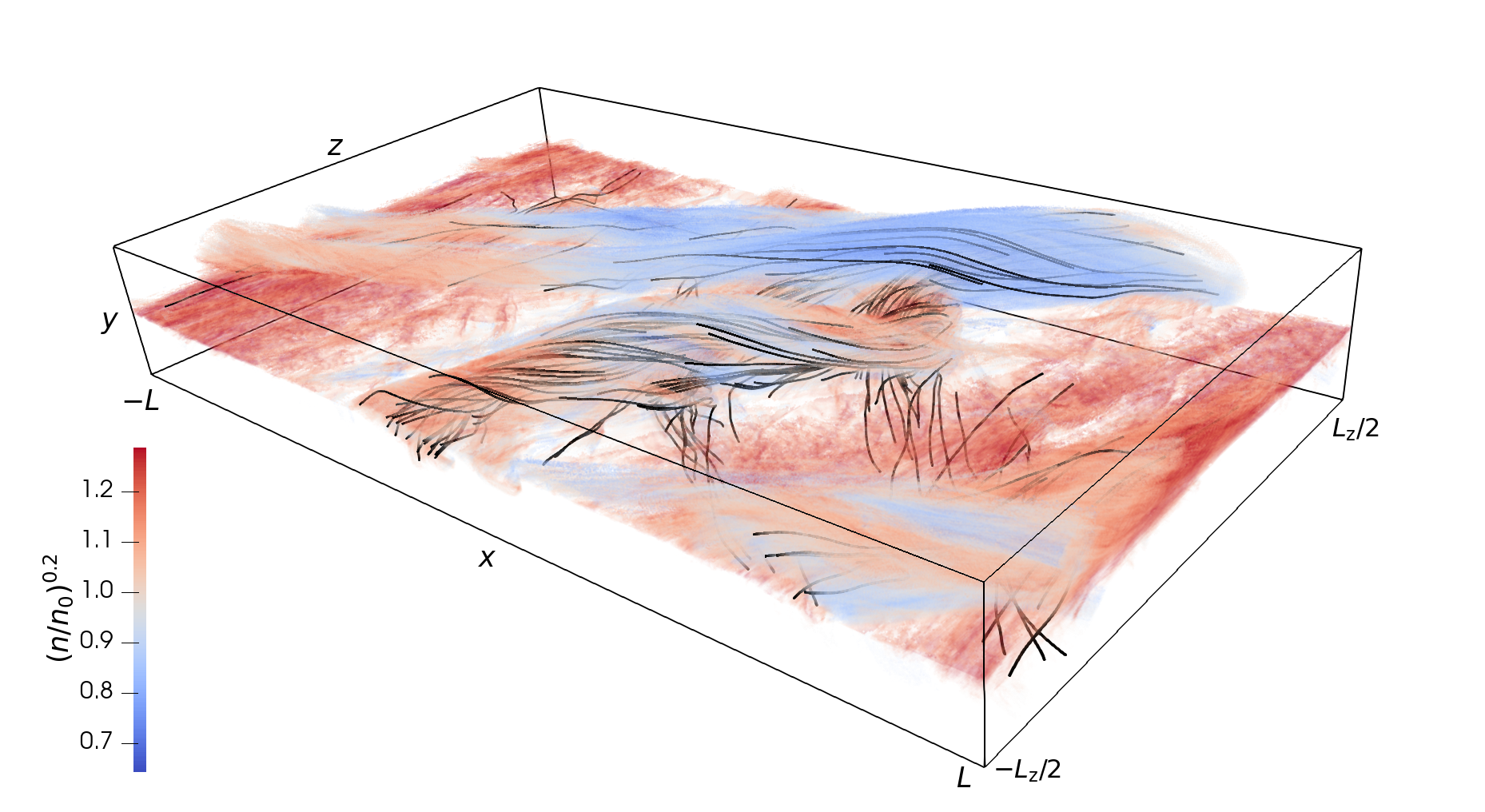}
    \caption{Volume rendering of the reconnected (downstream) plasma for $B_{\rm g}=0.0$ (left) and $B_{\rm g}=0.5\,B_{\rm 0}$ (right), identified using the mixing criterion described in the text and taken at $t=3.38 \,L/c$. Color indicates the plasma density normalized to its initial upstream value and raised to the power of $1/5$ \dk{for better visualization}. Solid black curves trace magnetic field lines.}
    \label{fig:current_sheets}
\end{figure*}

\begin{figure*}
    \centering
    \includegraphics[width=0.98\linewidth]{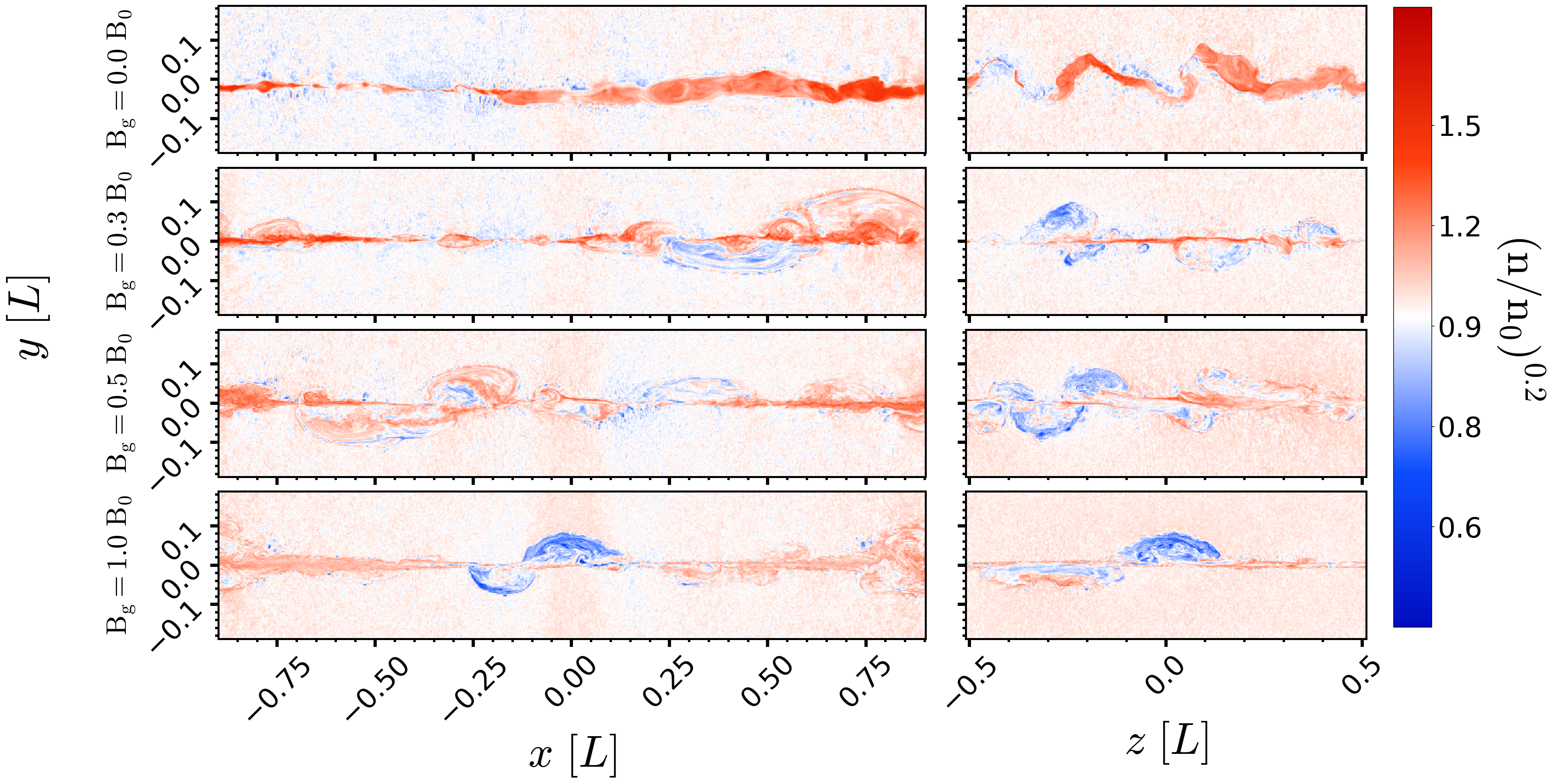}
    \caption{2D slices of the reconnection volume at $t=3.38\,L/c$ for different guide-field cases. The slices in the left column are taken in the $y$--$x$ plane at $z=0$, while those in the right column are taken in the $y$--$z$ plane at $x=0$ (we take $(x,y,z)=(0,0,0)$ at the center of the simulation box). Color indicates the plasma density normalized to its upstream value and raised to the power of $1/5$, as in  Figure \ref{fig:current_sheets}.}
    \label{fig:2D_slices}
\end{figure*}

For all simulations, we set the domain size along $x$ to $L_{\rm x}=4000$ cells and define $L\equiv L_{\rm x}/2$ as our characteristic length scale. In the 3D simulation, the domain size along $z$ is $L_{\rm z}=2040$ cells. Reconnection is initiated by locally reducing the particle pressure in the center of the $x$--$y$ plane. This perturbation triggers reconnection and launches two fronts that propagate in opposite directions along $x$, evacuating the plasma initially contained within the current layer after a time $\sim L/c$. 

We track a subset of $10^5$ particles in each simulation. Particles are randomly sampled from the simulation and retained in the tracked dataset if their Lorentz factor reaches $\gamma \geq \sigma$ at least once during their evolution. For each selected particle, we store the full trajectory with an output cadence of $18\,\omega_{\rm p}^{-1}$ \dk{for both 2D and 3D}, including timesteps preceding the threshold crossing. A particle is considered to have escaped, and is therefore no longer tracked, once it comes within 200 cells of either domain boundary along the $x$ direction, consistent with the discussion above. 

\section{Results} \label{Sec:results}

\subsection{Structure of the layer}
 \label{subsec:structure}
In order to visualize the structure of the reconnected (downstream) plasma, we show volume renderings for $B_{\rm g}=0.0$ and $0.5\,B_{\rm 0}$ in the left and right panels of Figure~\ref{fig:current_sheets}, respectively. The color scale represents the plasma density normalized to its initial upstream value and raised to the power of $1/5$. To distinguish the downstream from the upstream, we adopt the mixing criterion
\begin{equation}
    \mathcal{M}=1-2\left|\frac{n_{\uparrow}}{n}-\frac{1}{2}\right|,
\end{equation}
as defined by \citet{zhang_sironi_21}, where $n_{\uparrow}$ is the local number density of particles initialized at $y>0$ and $n$ is the total number density. The upstream corresponds to unmixed plasma, with $\mathcal{M}\approx 0$, whereas the downstream is strongly mixed, with $\mathcal{M}\approx 1$. We adopt a threshold of $\mathcal{M}\geq 0.3$ to identify the reconnected (downstream) plasma \citep{zhang_sironi_21}.

For zero guide field (left column), the downstream region is strongly corrugated along the $z$ direction, as a consequence of kink-like modes. The resulting density structures extend predominantly along the $x$ direction, likely due to stretching by the bidirectional reconnection outflows. These $x$-elongated structures are intermittently connected by flux tubes whose axes are approximately aligned with the $z$ direction, with magnetic field lines wound toroidally around them, as shown by the black curves in the figure.
In contrast, for a stronger guide field, $B_{\rm g}=0.5\,B_{\rm 0}$ (right column), the flux tubes are preferentially aligned with the upstream magnetic field, implying that their orientation reverses between the two sides of the layer. They also appear more ordered and coherent than in the zero-guide-field case, consistent with a suppression of kink-like instabilities as the guide-field strength increases. Their number decreases, while their characteristic cross section increases relative to the vanishing-guide-field case.

In addition to the volume renderings of Figure~\ref{fig:current_sheets} above, \dk{in Figure~\ref{fig:2D_slices} we present} 2D slices of the reconnection volume in the $y$--$x$ and $y$--$z$ planes, shown in the left and right columns respectively, for guide-field strengths $B_{\rm g}/B_{\rm 0}=0$, 0.3, 0.5, and 1.0 (the $B_{\rm g}/B_{\rm 0}=0.1$ case is statistically similar to the vanishing-guide-field case and is therefore omitted). These slices confirm the main conclusions inferred from the 3D renderings. For the vanishing-guide-field case (top row), the reconnection layer is strongly corrugated along the $z$ direction, and most of the  downstream structures correspond to regions of enhanced density. As the guide-field strength increases, the reconnected plasma becomes progressively more ordered, with the denser structures preferentially concentrated near the midplane. The density enhancement of the downstream plasma generally decreases with increasing guide-field strength, as the guide-field pressure inhibits compression. Unlike in the $B_{\rm g}=0$ case, regions both denser and less dense than the upstream plasma emerge; the underdense regions (blue) are particularly evident within the larger flux tubes.

\begin{figure*}
\centering
    \includegraphics[width=0.9\textwidth]{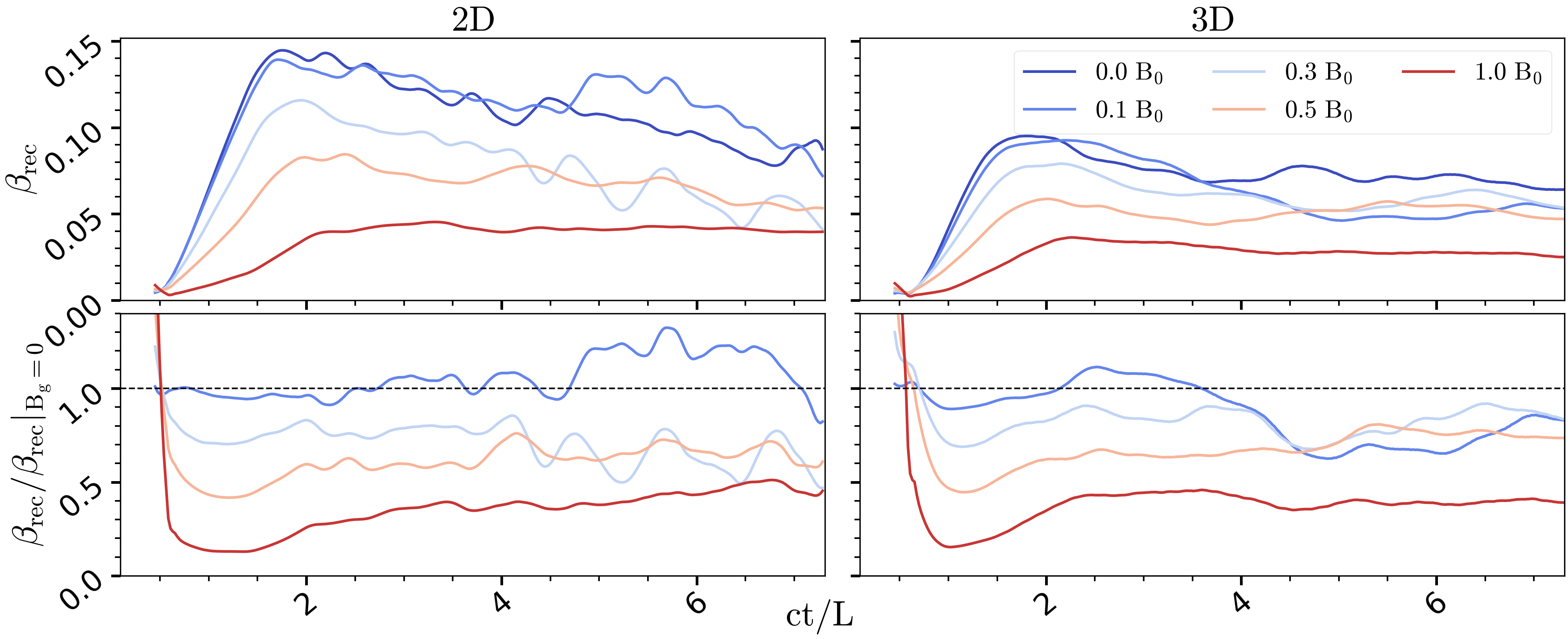}
    \caption{\textit{Top panels:} Time evolution of the reconnection rate, $\beta_{\rm rec} \equiv v_{\rm in}/c$, where $v_{\rm in}$ is the spatially averaged inflow speed, for the 2D (left) and 3D (right) simulations. \textit{Bottom panels:} Reconnection rate normalized to the instantaneous value of the corresponding $B_{\rm g}=0$ simulation. In all panels, color indicates the guide-field strength (see legend).}
    \label{fig:recrate}
\end{figure*}

\subsection{Reconnection Rate} \label{subsec:RecRate}
During reconnection, the upstream plasma is advected toward the layer. The dimensionless ratio of the inflow velocity to the speed of light,
$\beta_{\rm rec} \equiv v_{\rm in}/c$,
defines the reconnection rate. We measure the inflow speed as the spatially averaged velocity in the upstream, selecting a region sufficiently far from the reconnection layer to avoid contamination from plasmoid / flux-rope motions.

Figure~\ref{fig:recrate} presents the time evolution of $\beta_{\rm rec}$ for the 2D and 3D simulations (top left and top right panels, respectively) and for all guide-field strengths (see legend). The bottom panels show, for each nonzero guide field, the ratio of the reconnection rate to the instantaneous value of the corresponding zero-guide-field simulation.

In all cases, the initial increase in the reconnection rate reflects the growing volume undergoing active reconnection, as the two reconnection fronts generated by the initial perturbation at  $(x,y)=(0,0)$ propagate in opposite directions toward the $x$ boundaries. After this initial transient, most 3D runs reach a quasi-steady state by $t\sim 3\,L/c$.\footnote{The $B_{\rm g}=0.1\,B_{\rm 0}$ case is an exception: its reconnection rate continues to decrease until $t\sim 5\,L/c$, after which it stabilizes.} In the 2D simulations, the low-guide-field cases, up to $B_{\rm g}=0.3\,B_{\rm 0}$, decline slowly throughout the simulation, whereas the stronger-guide-field cases, $B_{\rm g}=0.5\,B_{\rm 0}$ and $B_{\rm g}=1.0\,B_{\rm 0}$, reach a quasi-steady state already by $t\sim 2\,L/c$. In what follows, we measure time-averaged properties over the quasi-steady interval $t\geq 3\,L/c$ for all cases.

Comparing the top left and top right panels, the reconnection rate in the 3D simulations is systematically lower than in the corresponding 2D cases for all guide-field strengths. The temporal evolution in 3D is smoother than in 2D, because the additional averaging along the $z$ direction makes the measured rate less sensitive to variability in the layer dynamics. Despite the systematic reduction in $\beta_{\rm rec}$ from 2D to 3D, the bottom panels show that increasing $B_{\rm g}$ reduces $\beta_{\rm rec}$ by nearly the same factor in both geometries. The only possible exception is the $B_{\rm g}=0.1\,B_{\rm 0}$ case: at late times, the 2D reconnection rate becomes slightly larger than in the zero-guide-field run, whereas in 3D it remains slightly smaller. Overall, in agreement with earlier results \citep[e.g.,][]{werner_17}, we find that stronger guide fields progressively reduce the reconnection rate, by a comparable relative amount in both 2D and 3D simulations.

\begin{figure}
    \centering
    \includegraphics[width=0.95\columnwidth]{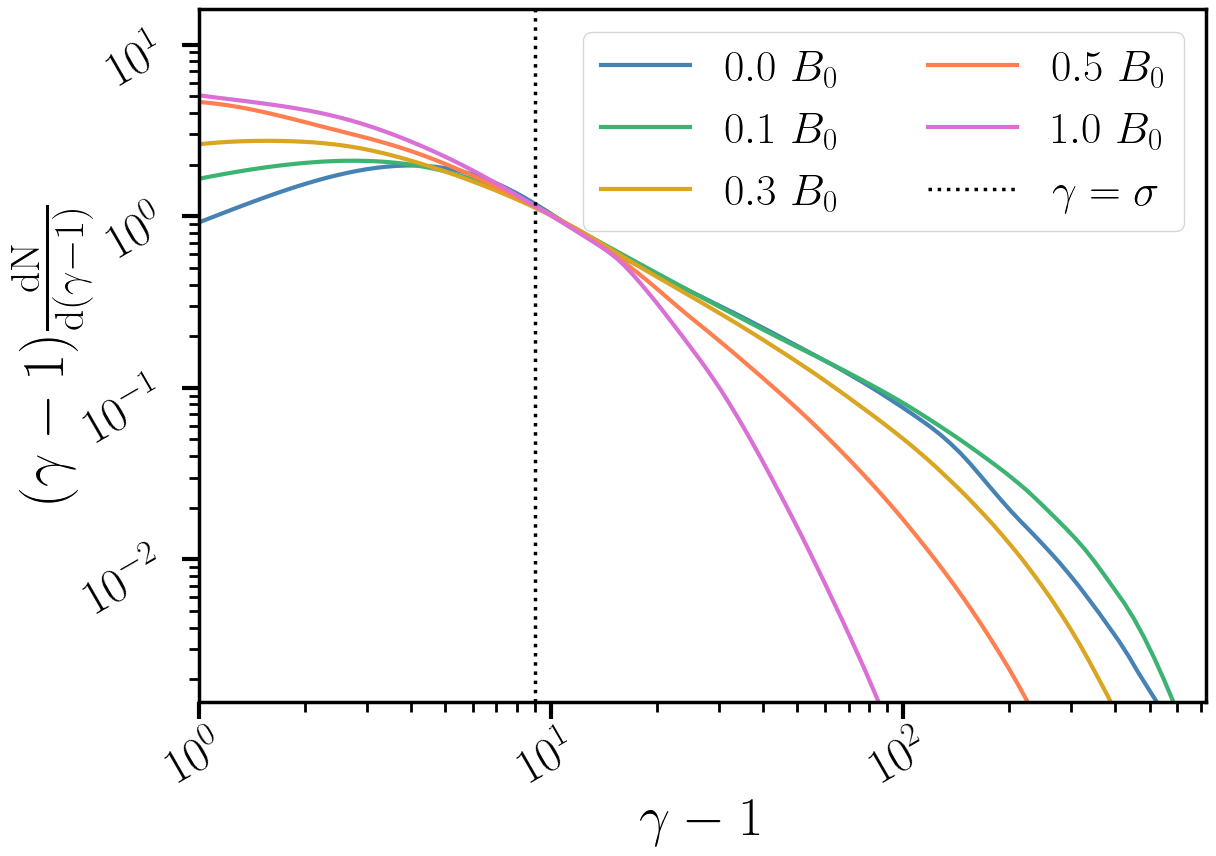}
    \caption{Time-averaged energy spectra for $t\geq 3\,L/c$, extracted from the 3D simulations for all guide-field strengths (see legend). Each spectrum is normalized to its value at $\gamma=\sigma$, indicated by the vertical dotted line.}
    \label{fig:spec_3D}
\end{figure}

\begin{figure}
    \centering
    \includegraphics[width=0.99\columnwidth]{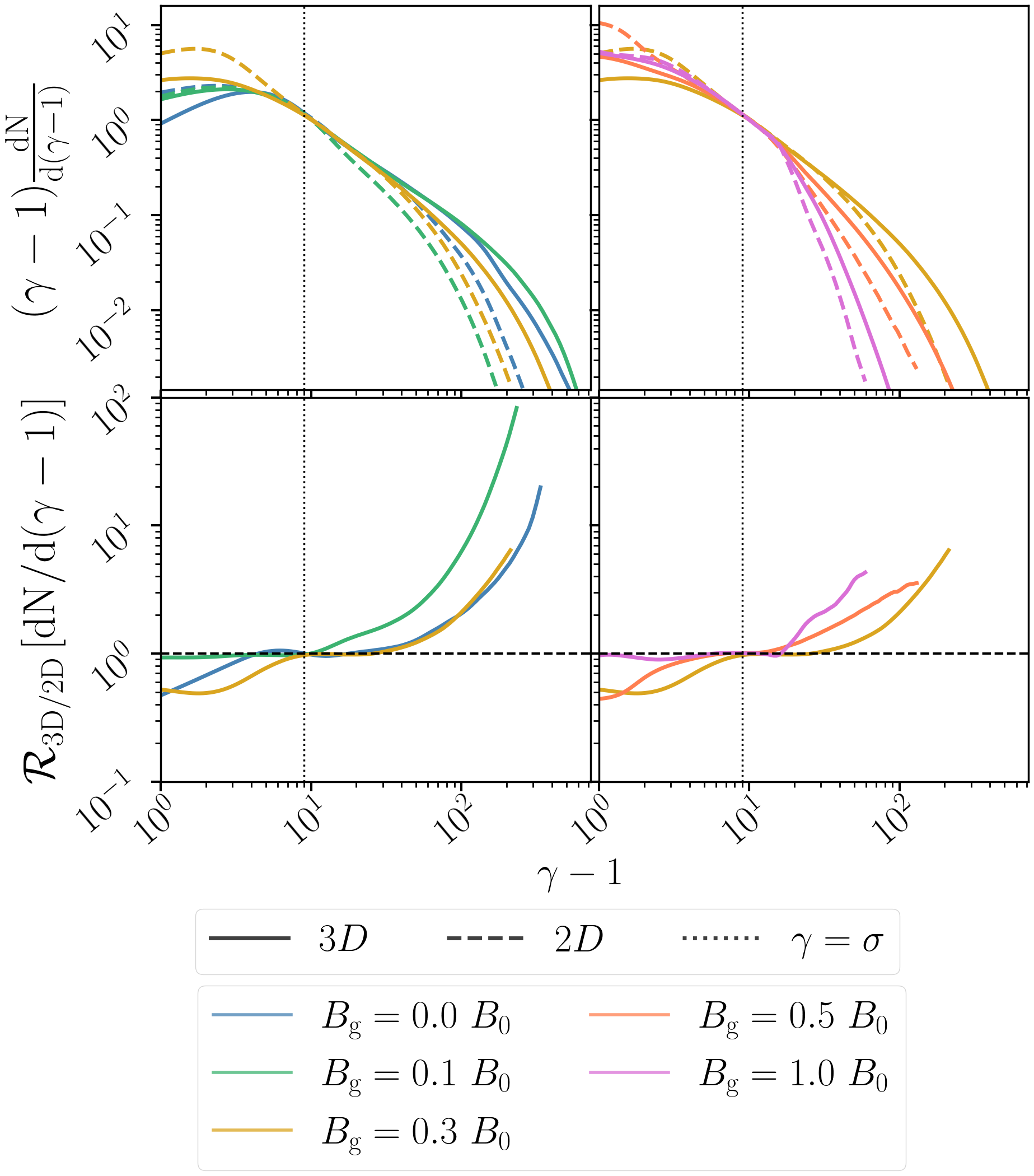}
    \caption{\textit{Top panels:} Time-averaged energy spectra for $t\geq 3\,L/c$ for all guide-field strengths. Solid and dashed colored lines show the 3D and 2D simulations, respectively. \textit{Bottom panels:} Ratio $\mathcal{R}$ of the 3D particle spectrum to the corresponding 2D spectrum. The left panels show the lower-guide-field cases, $B_{\rm g}/B_{\rm 0}=0$, 0.1, and 0.3, while the right panels show the higher-guide-field cases, $B_{\rm g}/B_{\rm 0}=0.3$, 0.5, and 1.0. Each spectrum is normalized to its value at $\gamma=\sigma$, indicated by the vertical dotted line.}
    \label{fig:spec_3Dvs2D}
\end{figure}

\subsection{Particle Energy Spectra}
In this section, we present the time-averaged positron energy spectra, averaged over the interval $t\geq 3\,L/c$, when the system has reached a quasi-steady state (see Figure~\ref{fig:recrate}), and normalized to their value at $\gamma=\sigma$ to facilitate visual comparison. The corresponding electron spectra are nearly indistinguishable. Throughout this section, we consider the combined contribution of particles residing in the upstream and downstream regions.

Previous 3D studies focusing on weak guide fields \citep[e.g.][]{zhang_sironi_21, zhang_23} used a threshold of $\mathcal{M}=0.3$ to distinguish between the upstream and downstream regions. This allowed the particle spectra to be separated into ``free'' particles residing in the upstream and ``trapped'' particles residing in the downstream. We find, however, that the mixing criterion becomes progressively less reliable for distinguishing between these two regions as the guide-field strength increases. Developing a more robust criterion for separating the free and trapped populations will be left to future work.

Figure~\ref{fig:spec_3D} shows the overall particle spectra of the 3D simulations, including contributions from both upstream and downstream, for all guide-field strengths considered (see legend). As the guide field increases, the spectra become progressively steeper and extend to lower maximum Lorentz factors, in agreement with earlier studies \citep[e.g.,][]{werner_17}. Below, we show that the fast particle-acceleration channel observed in 3D simulations with weak guide fields \citep{zhang_sironi_21, zhang_23}, and associated there with the free--trapped sequence, continues to operate even for strong guide fields. This suggests that the spectral steepening observed at larger $B_{\rm g}$ may result, at least in part, from a progressively smaller fraction of particles participating in the fast, intrinsically 3D acceleration process. We test this interpretation in the next subsection.

The top panels of Figure~\ref{fig:spec_3Dvs2D} present the time-averaged particle energy spectra for $t\geq 3\,L/c$, comparing the 3D simulations (solid colored lines) with their 2D counterparts (dashed colored lines) for all guide-field strengths considered (see legend). All spectra are normalized to their value at $\gamma=\sigma$, indicated by the vertical dotted line. The bottom panels show the ratio of the 3D to 2D spectra for each guide-field case. The figure demonstrates that increasing $B_{\rm g}$ suppresses particle acceleration in both 2D and 3D, with particles reaching progressively lower maximum Lorentz factors as the guide-field strength increases (see also Figure~\ref{fig:spec_3D}). In addition, comparison of the solid and dashed curves of the same color in the top panels shows that 2D simulations systematically underpredict the high-energy tail, $\gamma\gtrsim\sigma$, relative to their 3D counterparts. The difference is particularly pronounced for $B_{\rm g}=0.1\,B_{\rm 0}$, but persists across the full range of guide-field strengths explored.

\begin{figure}
    \centering
    \includegraphics[width=0.99\linewidth]{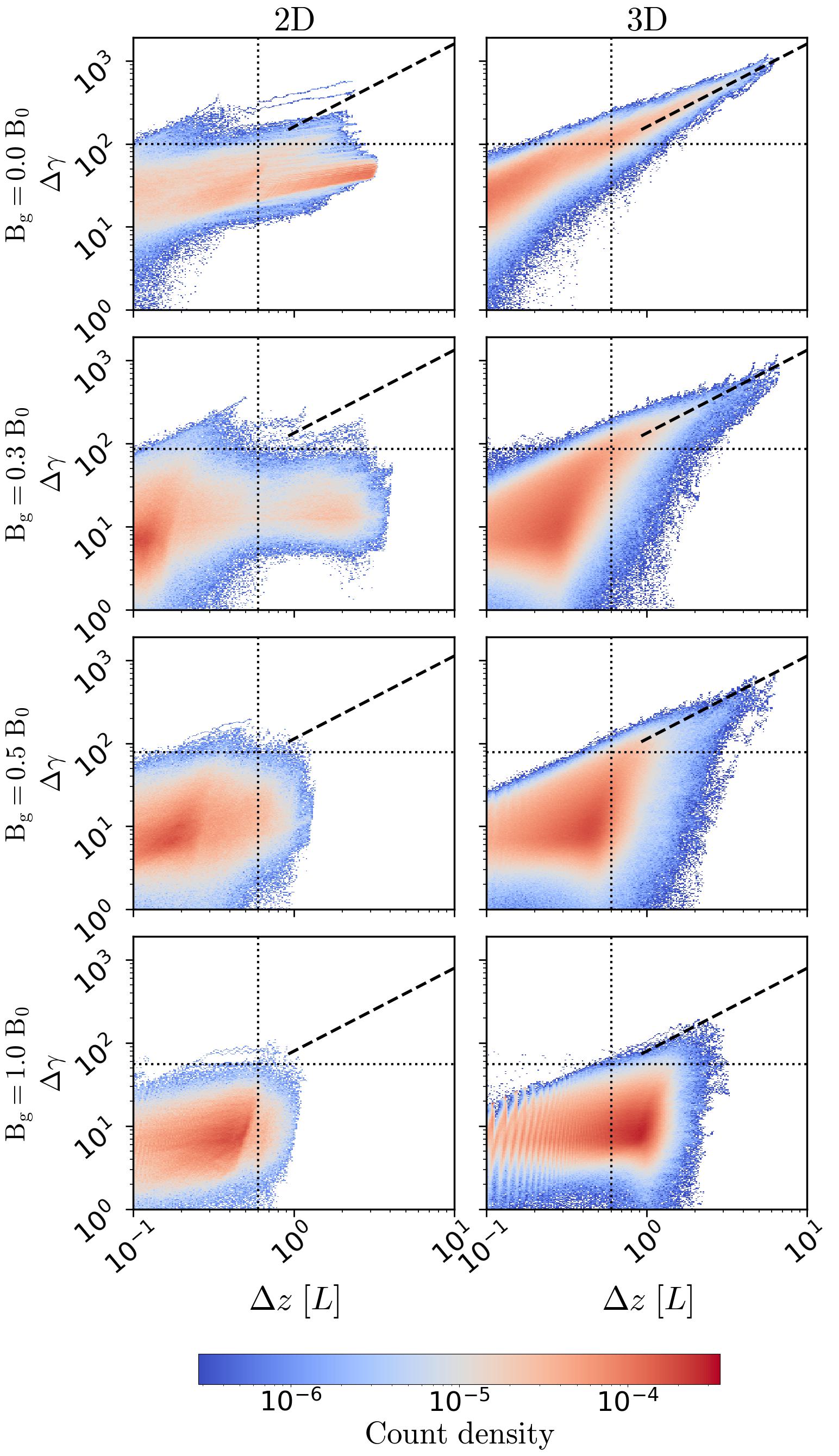}
    \caption{2D histograms for the $10^5$ positrons tracked in each simulation. The vertical axis shows the change in positron Lorentz factor relative to its initial value, while the horizontal axis shows the displacement along the $z$ direction. The left and right columns correspond to the 2D and 3D simulations, respectively, while different rows show different guide-field strengths. Horizontal and vertical dotted lines mark the thresholds in $\Delta\gamma$ and $\Delta z$, respectively, with $\Delta z_{\rm th}=0.6\,L$ and $\Delta\gamma_{\rm th}$ given by Eq.~\ref{eq:Dg_linear} for $\alpha=10$.}
    \label{fig:Dg_Dz_hist}
\end{figure}

\subsection{Particle Acceleration Rate}\label{sec:acceleration}
As discussed in Section \ref{Sec:Intro}, 3D PIC simulations of relativistic reconnection in the weak-guide-field regime have revealed a fast particle-acceleration channel that is absent in 2D \citep{zhang_sironi_21,zhang_23}. For weak guide fields, this mechanism is associated with a sequence of free and trapped phases and is characterized by an acceleration rate that can approach the theoretical maximum, $\dot{\gamma}_{\rm acc} \sim \beta_{\rm rec}\,\omega_{\rm c}$. In this subsection, we investigate whether evidence for acceleration near this maximal rate persists for moderate and strong guide fields.

In Figure~\ref{fig:Dg_Dz_hist}, we present two-dimensional histograms that help characterize the physics of particle acceleration for the $10^5$ tracked positrons described in Section~\ref{sec:Setup} (randomly sampled as long as their Lorentz factor exceeds $\sigma$ at least once during their evolution). The vertical axis shows the change in positron Lorentz factor relative to its initial value, $\Delta\gamma=\gamma-\gamma_{\rm 0}$, while the horizontal axis shows its displacement along the $z$ direction, $\Delta z=z-z_{\rm 0}$. Color indicates the count density in each bin, with each particle contributing to the histogram once for every saved output.
The left and right columns correspond to 2D and 3D simulations, respectively, while the rows show results for different guide fields (increasing from top to bottom).
\dk{Regarding the 3D aspects of fast particle acceleration, the $B_{\rm g}/B_0 = 0.1$ case closely resembles the case with vanishing guide field, although it exhibits a slightly higher reconnection rate and (consequently) a higher spectral cutoff Lorentz factor. Since a similar analysis for the $B_{\rm g}/B_0 = 0.1$ case has been presented in \citet{zhang_sironi_21, zhang_23}, we do not repeat it here.}
In the 3D simulations, the $z$ coordinate is recorded directly by the code. In the 2D simulations, instead, the $z$ displacement is reconstructed from the $z$ component of the particle velocity, which is saved with an output cadence of $18\,\omega_{\rm p}^{-1}$ , by integrating in time under the assumption that the velocity remains constant between consecutive outputs.\footnote{The nearly vertical stripes visible in the bottom-right panel at $\Delta z\lesssim 0.5\,L$ are a consequence of the finite output cadence. For a particle moving along the $z$ direction at the speed of light, the cadence of $18\,\omega_{\rm p}^{-1}$ implies discrete displacement increments of $18\,c/\omega_{\rm p}$, producing vertical bands at approximately integer multiples of this value. \dk{This feature is stronger in the 3D case of $B_{\rm g}/B_{\rm 0}=1.0$ since in that case the particles move preferentially along the $z$ direction}.}

For $B_{\rm g}/B_{\rm 0}=0$ and 0.3, the histograms of the 2D simulations (left column, top two rows) exhibit a prominent overdensity in the bottom-right quadrant, at $10\lesssim\Delta\gamma\lesssim100$ and $\Delta z/L\gtrsim1$ (the choice of the limits that define the quadrants, shown as vertical and horizontal lines, is explained below). We have verified that the particles contributing to this feature are localized near the cores of large plasmoids, where they undergo strong drifts along the $z$ direction, driven by magnetic-field gradients and curvature, without experiencing significant energization. This feature is absent in the corresponding 3D simulations (right column).

The most striking difference between 2D and 3D simulations in Figure~\ref{fig:Dg_Dz_hist} is the emergence, in 3D, of a distinct population of particles occupying the upper-right quadrant, which are absent in 2D. These particles exhibit a positive correlation between energy gain and displacement along the $z$ direction, $\Delta\gamma\propto\Delta z$. This component, earlier identified in 3D simulations with weak guide fields \citep{zhang_sironi_21, zhang_23}, is shown to persist across the full range of guide-field strengths investigated here. The positive correlation $\Delta\gamma\propto\Delta z$ is expected for particles accelerated by the reconnection electric field, $E_{\rm rec}\sim\beta_{\rm rec}B_{\rm 0}$, while moving along the $z$ direction at nearly the speed of light. For these fast-accelerated particles, we expect
\begin{equation}
    \frac{\Delta\gamma}{\Delta z}
    \sim
    \frac{eE_{\rm rec}}{m_{\rm e}c^2}
    \sim
    \frac{\beta_{\rm rec}\sqrt{\sigma}\,\omega_{\rm p}}{c}~.
    \label{Eq:Dgdz}
\end{equation}
which is indicated by the dashed black lines in 
Figure~\ref{fig:Dg_Dz_hist}, taking $\beta_{\rm rec}$ as the time-averaged reconnection rate.

We now quantify the fraction of particles undergoing this fast acceleration process by identifying the upper-right quadrant through thresholds in $\Delta z$ and $\Delta\gamma$. We adopt a fixed displacement threshold, $\Delta z_{\rm th}=0.6\,L$, for all guide-field strengths, and define a guide-field-dependent energy threshold as
\begin{equation}
    \Delta \gamma_{\rm th}
    = \alpha \sigma
    \frac{\langle \beta_{\rm rec}(B_{\rm g}) \rangle}
    {\langle \beta_{\rm rec}(B_{\rm g}=0) \rangle},
    \label{eq:Dg_linear}
\end{equation}
where $\langle \beta_{\rm rec}\rangle$ is the reconnection rate averaged over the quasi-steady-state interval ($t\geq 3\,L/c$). Its dependence on $B_{\rm g}$ in the 3D simulations is well approximated by the linear fit
\begin{equation}
    \langle \beta_{\rm rec} \rangle
    \simeq
    -0.025^{\pm 0.009}(B_{\rm g}/B_{\rm 0})
    + 0.057^{\pm 0.004}.
    \label{eq:beta_fit}
\end{equation}
Comparison of Eq.~\ref{eq:Dg_linear} and Eq.~\ref{Eq:Dgdz} allows to infer that $\alpha=0.6 \,(L/r_{\rm L,hot})\langle \beta_{\rm rec}(B_{\rm g}=0) \rangle$, where $r_{\rm L,hot}=\sqrt{\sigma}\,c/\omega_{\rm p}=\sigma\, c/\omega_{\rm c}$ is the characteristic Larmor radius of particles heated by reconnection. For our simulations, $\alpha\sim 10$. The upper-right quadrant in each panel, defined by $\Delta z>\Delta z_{\rm th}$ and $\Delta\gamma>\Delta\gamma_{\rm th}$ (dotted vertical and horizontal lines, respectively), is the region in which particles are undergoing fast acceleration. Figure~\ref{fig:Dg_Dz_hist} shows that this region is substantially more populated in 3D than in the corresponding 2D simulations. However, for stronger guide fields it becomes progressively less populated, suggesting that the fast acceleration channel is progressively suppressed as the guide field increases.

\begin{figure}
    \centering
    \includegraphics[width=0.99\linewidth]{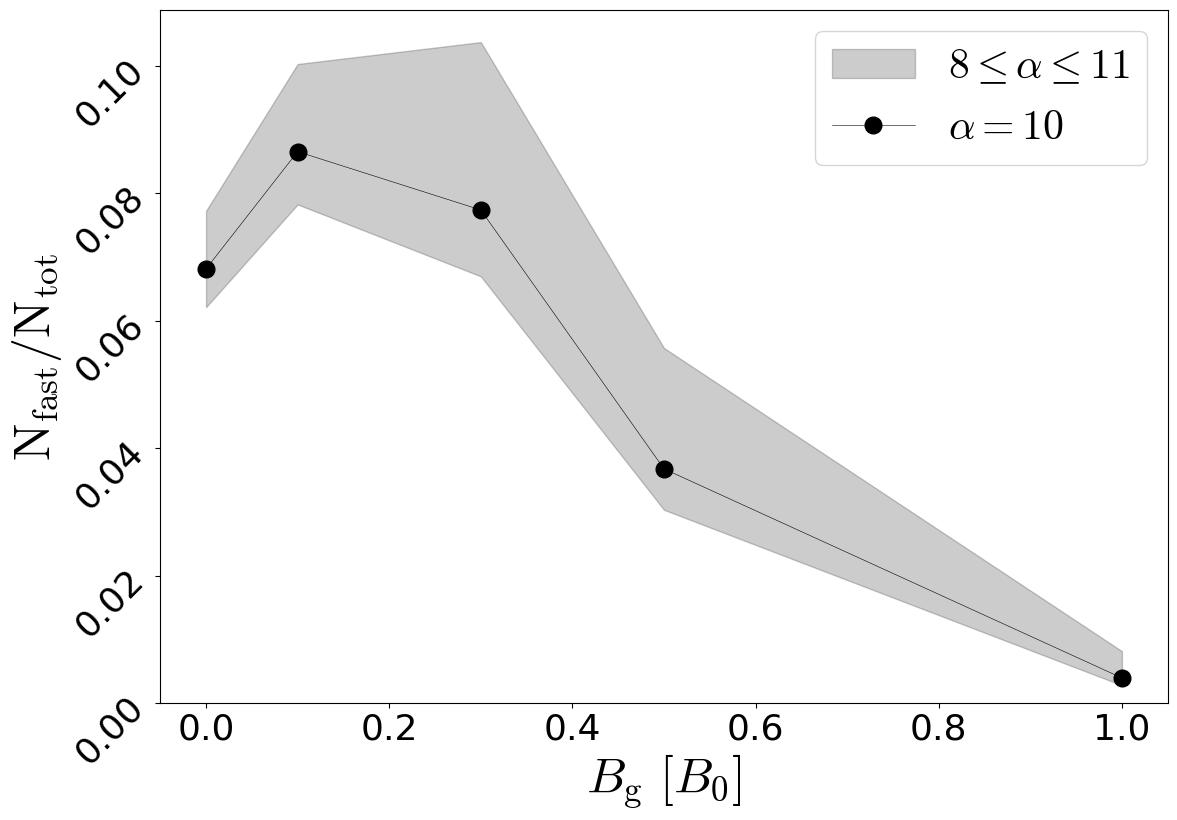}
    \caption{Black markers show the fraction of particles undergoing fast acceleration, defined as those entering the upper-right quadrant of Figure~\ref{fig:Dg_Dz_hist} for $\alpha=10$ in Eq.~\ref{eq:Dg_linear}. The gray-shaded region indicates the variation in this fraction obtained by varying $\alpha$ over the range $8\leq\alpha\leq11$.}
    \label{fig:3D_perc}
\end{figure}

We identify as fast-accelerating those particles that, at some point during their evolution, enter the upper-right quadrant defined by $\Delta z>\Delta z_{\rm th}$ and $\Delta\gamma>\Delta\gamma_{\rm th}$. The fraction of such particles relative to the total tracked sample ($10^5$ particles for each simulation) is evaluated as a function of guide-field strength and shown by the black markers in Figure~\ref{fig:3D_perc}, adopting $\alpha=10$ in Eq.~\ref{eq:Dg_linear}. This fraction is largest at weak guide fields and decreases substantially as $B_{\rm g}$ increases, from $\simeq 0.07$ for vanishing guide field to $\simeq 0.005$ for $B_{\rm g}=B_{\rm 0}$. This trend mirrors the progressive depletion of the upper-right quadrant seen in Figure~\ref{fig:Dg_Dz_hist}, indicating that fewer particles participate in the fast acceleration channel at stronger guide fields. Interestingly, the fraction appears to peak at $B_{\rm g}=0.1\,B_{\rm 0}$, the guide-field strength adopted by \citet{zhang_sironi_21, zhang_23}. While the overall trend with $B_{\rm g}$ is insensitive to the choice of $\alpha$, the precise fraction of fast-accelerating particles does depend on this parameter, as illustrated by the grey-shaded region in Figure~\ref{fig:3D_perc}, where $\alpha $ is varied between 8 and 11, with lower values yielding higher fractions.

\begin{figure}
    \centering
    \includegraphics[width=0.99\linewidth]{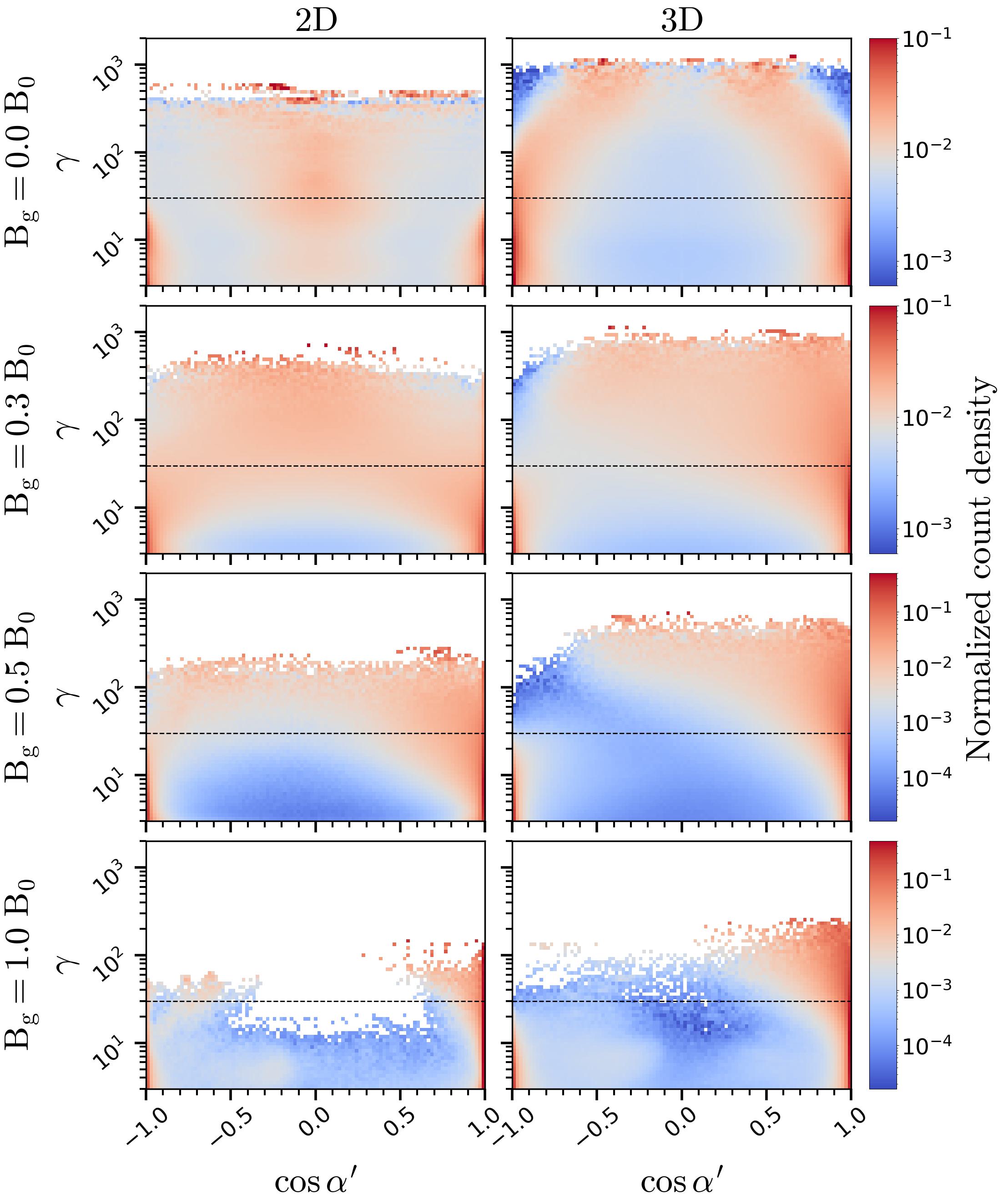}
    \caption{2D histograms of the positron Lorentz factor, measured in the simulation frame, versus the cosine of the pitch angle, measured in the local drift frame moving with $c\,\mathbf{E}\times\mathbf{B}/B^2$, for the 2D (left column) and 3D (right column) simulations. Different rows correspond to increasing guide-field strengths from top to bottom. Each positron contributes to the histogram once for every saved output (during the quasi-steady interval) in which $\gamma>3$. Color indicates the number of counts, normalized by the total number of counts in each row (i.e., in each $\gamma$-bin). The horizontal dotted lines mark $\gamma=3\sigma$, the threshold identified by \citet{zhang_sironi_21} for the free--trapped sequence associated with fast particle acceleration.}
    \label{fig:contours}
\end{figure}

\subsection{Pitch Angle Distribution}
In this subsection, we analyze the pitch-angle distribution of all positrons with $\gamma>3$ during the quasi-steady interval. The pitch angle is computed in the local frame defined by the drift velocity $ c \mathbf{E}\times\mathbf{B}/{B^2}$, where the electromagnetic fields $\mathbf{E}$ and $\mathbf{B}$ are evaluated at the instantaneous particle position. The pitch angle, $\alpha^{\prime}$, is then defined in the drift frame as the angle between the particle momentum and the local magnetic field.

Figure~\ref{fig:contours} shows two-dimensional histograms of the positron Lorentz factor, measured in the simulation frame, versus $\cos\alpha^{\prime}$, measured in the drift frame. \dk{We retain the simulation-frame $\gamma$ because it is only weakly affected by
Lorentz transformations, unlike the pitch-angle cosine, which can differ
substantially between the two frames and, for computing synchrotron emission, should
be evaluated in the drift frame. Retaining the simulation-frame $\gamma$ also
facilitates direct comparison with our earlier figures.} The left and right columns correspond to the 2D and 3D simulations, respectively, while the rows from top to bottom refer to increasing guide-field strength. Color indicates the count density, defined as the number of counts in each bin normalized by the total number of counts in the corresponding $\gamma$-bin (i.e., in each row). We have checked that the corresponding electron distribution is mirror-symmetric along the horizontal axis.

The pitch-angle distributions exhibit several trends that are common to both 2D and 3D simulations, as well as important differences between the two geometries. For all guide-field strengths and in both dimensionalities, low-energy particles with $\gamma\lesssim\sigma$ are preferentially aligned with the local magnetic field, $|\cos\alpha^{\prime}|\simeq 1$ \citep[as previously reported in 2D studies by][]{comisso_pitch-angle_2023, comisso_concurrent_2024}, whereas higher-energy particles span a broader range of pitch angles and progressively shift toward smaller values of $|\cos\alpha^{\prime}|$, corresponding to momenta more nearly perpendicular to the local magnetic field.  This transition toward small $|\cos\alpha^{\prime}|$ at high energies is more pronounced for weaker guide fields.

For a vanishing guide field (top row), the distributions are approximately symmetric about $\cos\alpha^{\prime}=0$, in both 2D and 3D. \dk{As the guide-field strength increases, the distributions become increasingly peaked at $|\cos\alpha^{\prime}|\simeq 1$, with some tendency for positrons to accumulate near $\cos\alpha^{\prime}\simeq 1$.}
In our geometry, this corresponds to motion aligned with the local magnetic field and, increasingly for stronger guide fields, with the direction of the reconnection electric field in the simulation frame. \dk{The marginal preference for $\cos\alpha^{\prime}\simeq 1$ as compared to $\cos\alpha^{\prime}\simeq -1$} is therefore controlled by the direction of the reconnection electric field. Consistent with this interpretation, we have checked that the corresponding two-dimensional histograms for electrons show the opposite asymmetry, with most low-energy electrons concentrated near $\cos\alpha^{\prime}\simeq -1$. 

The most pronounced difference between 2D and 3D simulations at fixed guide field strength appears at high energies, $\gamma\gtrsim3\sigma$. For weak guide fields, the 2D simulations show a stronger concentration of high-energy particles near $\cos\alpha^{\prime}=0$. This likely reflects the dominant acceleration mechanism for high-energy particles in 2D, where particles remain confined near plasmoid cores and gain energy as the magnetic-field strength increases, approximately conserving their magnetic moment \citep{petropoulou_18,hakobyan_21}. This  increases the perpendicular momentum relative to the parallel momentum, driving $\cos\alpha^{\prime}$ toward zero.

In 3D, by contrast, the transition toward $\cos\alpha^{\prime}=0$ at high particle energies is more gradual. For weak guide fields, this transition likely reflects the increasing importance of freely accelerating particles relative to trapped particles at higher energies. In the picture presented by \citet{zhang_sironi_21}, particles in the free phase move predominantly perpendicular to the upstream magnetic field, corresponding to $\cos\alpha^{\prime}\simeq0$. As their relative contribution increases toward the high-energy end of the distribution, they progressively shift the overall pitch-angle distribution toward smaller $|\cos\alpha^{\prime}|$. This trend is present at all guide-field strengths, but becomes progressively less pronounced as $B_{\rm g}$ increases, consistent with the results of the previous subsection showing that the fraction of particles participating in the fast acceleration channel decreases with increasing guide-field strength.

\section{Summary and Discussion} \label{Sec:summary}
In this work, we have performed 3D PIC simulations of relativistic pair-plasma reconnection, employing open boundaries along the reconnection outflows to establish a quasi-steady state. In agreement with earlier results obtained in periodic domains \citep[e.g.,][]{werner_17}, we find that increasing the guide-field strength reduces the reconnection rate. 

Our analysis focuses primarily on how the guide field affects particle acceleration. We find that the fast acceleration mechanism previously identified in 3D simulations with weak guide fields, and absent in 2D \citep{zhang_sironi_21,zhang_23}, persists across the full range of guide-field strengths investigated here. However, the fraction of particles participating in this channel decreases progressively as the guide field increases. Consequently, stronger guide fields produce steeper particle energy spectra and lower maximum Lorentz factors. We have also characterized the pitch-angle distributions of energetic particles, finding that \dk{they become increasingly anisotropic as the guide field increases, with positrons preferentially concentrated around $\cos\alpha^{\prime} \simeq 1$ (and electrons around $\cos\alpha^{\prime} \simeq -1$)}, corresponding to momenta increasingly \dk{parallel} to the local field.

These results have direct implications for models of reconnection-powered emission in pulsar and black hole magnetospheres, magnetar flares, and relativistic jets \citep[][and references therein]{sironi_review_25}. Stronger guide fields reduce the fraction of particles undergoing fast acceleration, leading to steeper spectra and lower maximum energies, although the fast particle acceleration channel identified by \citet{zhang_sironi_21} persists even for $B_{\rm g}\sim B_{\rm 0}$. Future work will provide a statistical characterization of particle orbits in strong guide fields and determine how they shape the resulting power-law  energy spectrum, extending the analysis of \citet{zhang_23} to strong guide fields. 

Since the particle pitch-angle distribution becomes increasingly anisotropic with guide-field strength, this potentially provides an additional observational diagnostic that complements measurements of the power-law slope and cutoff energy. In particular, this anisotropy may imprint itself on polarization signatures from reconnection-powered emission \citep[e.g.][]{zhang_15}. \dk{We defer a quantitative assessment of these polarization signatures to future work.}

\acknowledgments
L.S. acknowledges support from DoE Early Career Award DE-SC0023015, NASA ATP 80NSSC24K1238, NASA ATP 80NSSC24K1826, and NSF AST-2307202. This research was facilitated by the Multimessenger Plasma Physics Center (MPPC) NSF grant PHY-2206609 to L.S., and by a grant from the Simons Foundation (MP-SCMPS-0000147, to L.S.).





\bibliography{blob2.bib}

\end{document}